# Growth and Characterization of Iron Scandium Sulfide (FeSc$_2$S$_4$)

J.R. Morey[a,b], K.W. Plumb[b], C.M. Pasco[a,b], B.A. Trump[a,b], T.M. McQueen[a,b,c,*], S.M. Koohpayeh[b,*]

[a] Department of Chemistry, Johns Hopkins University, Baltimore, MD 21218, USA

[b] Institute for Quantum Matter and Department of Physics and Astronomy, Johns Hopkins University, Baltimore, MD 21218, USA

[c] Department of Materials Science and Engineering, Johns Hopkins University, Baltimore, MD 21218, USA

**Abstract**

Here we report successful growth of mm scale single crystals of stoichiometric FeSc$_2$S$_4$. Single crystal X-ray diffraction yields a cubic structure, spacegroup $Fd\bar{3}m$, with $a$ = 10.5097(2) Å at $T$ = 110(2) K consistent with previous literature on polycrystalline samples. Models fit to the data reveal no detectable antisite mixing or deviations from the ideal stoichiometry. Heat capacity and dc magnetization measurements on the single crystals match those of high quality powder specimens. The novel traveling solvent crystal growth method presented in this work opens the door to studies requiring sizable single crystals of the candidate spin-orbital liquid FeSc$_2$S$_4$.

## 1. Introduction

Iron scandium sulfide, a cubic spinel of the form $AB_2X_4$, Fig. 1a, is a material of great interest as a candidate spin-orbital liquid (SOL), and has been the subject of many experimental [1-7] and theoretical studies [8-10]. The nature of SOLs remains an elusive subject for experimental observation, and the synthesis of more candidate SOLs enables further study in this area. Unfortunately, progress in understanding and utilizing $FeSc_2S_4$ has been hampered by the lack of stoichiometric powders and single crystals.

Previously reported growths of single crystal $FeSc_2S_4$ are via the iodine chemical vapor transport method [11]. Unfortunately, crystals produced in this fashion are quite small, yielding typical crystals of volume ~6 $\mu m^3$ after 30 days of reaction, and are often not quite stoichiometric due to changes in Fe/Sc ratio and substitution of I for S [11]. Here we report the first growths of stoichiometric single crystals of $FeSc_2S_4$ by the travelling solvent technique [12, 13], with an FeS solvent, in an optical heating furnace. Further, we present a synthetic route to stoichiometric, polycrystalline $FeSc_2S_4$ that does not involve the use of toxic and explosive $H_2S$ gas.

## 2. Experimental Methods

### 2.1 Preparation of Polycrystalline $FeSc_2S_4$

Some of the main difficulties associated with the growth of this material are (1) the volatilization of sulfur during each stage of the precursor synthesis and growth process, and (2) the reactivity of precursor materials with quartz and alumina crucibles. Additionally, Fe, FeS, $Sc_2S_3$, and $FeSc_2S_4$ are all moisture sensitive, and thus were handled in an argon glovebox with $pO_2$ < 2 ppm and $pH_2O$ < 1 ppm.

FeS powder was synthesized by heating stoichiometric amounts of sulfur pieces (99.999%, metals basis, Alfa Aesar) and vacuum remelted iron (99.99%, low oxygen, Alfa Aesar) in an evacuated quartz ampoule. $Sc_2S_3$ powder was synthesized by heating scandium metal pieces (99.9%, distilled dendritic (REO), Alfa Aesar) and sulfur pieces (99.999%, metals basis, Alfa Aesar) in a boron nitride (BN) crucible in an evacuated quartz ampoule. BN was used to prevent the reaction of scandium with quartz.

Sulfur volatilization limits the size of the quartz ampoules used by this method, as gaseous sulfur exerts an enormous amount of pressure in the tube. Ultimately, tubes of size 12x16 mm (IDxOD) were found to work best for larger scale (1 g) syntheses, and 4x6 mm tubes for smaller batches. $Sc_2S_3$ is the most sulfur rich phase in this system, so excess sulfur may be added to account for volatilization during sealing of the ampoule.

To prepare $FeSc_2S_4$ powder, FeS and $Sc_2S_3$ in a 1:1 molar ratio were ground together, pressed into a pellet, sealed in an evacuated quartz ampoule, and double sealed into a larger evacuated quartz ampoule. This was quickly heated to 500 °C, held for 2h, then heated at a rate of 50 °C/hr to 1000 °C. The furnace was allowed to cool to room temperature after 30 hours of reaction at 1000 °C.

Powder X-ray diffraction patterns were collected on a Bruker D8 Focus diffractometer with a LynxEye detector using Cu Kα radiation. Lattice parameters and Rietveld refinements were performed using Topas 4.2 (Bruker).

**2.2 Single crystal growth of $FeSc_2S_4$**

The pure polycrystalline $FeSc_2S_4$ powder and solvent (FeS) in a 2-3:1 mass ratio were placed in a pyrolytic boron nitride crucible with the solvent at the bottom of the container. The crucible was sealed in a quartz tube under 0.3 bar argon and placed vertically in an optical heating furnace (Crystal Systems Inc. FZ-T-4000-H-VII-VPO-PC). The focused radiation at 80-85% lamp power melted the solvent which was then moved along the polycrystalline $FeSc_2S_4$ powder at a traveling rate of 0.3 mm/h. Single crystals of mm size were grown by this technique.

**2.3 $FeSc_2S_4$ crystal characterization**

Single crystal X-ray diffraction data were collected using a SuperNova diffractometer equipped with an Atlas detector, irradiated with Mo Kα. The cuboid crystal, cut from a larger crystal piece, was mounted with Paratone-N oil. Diffraction patterns were analyzed using the CrysAlisPro software suite, version 1.171.36.32 (2013), Agilent Technologies. This software was also used to perform data reduction. Initial structural models were developed using SIR92 [14] and refinements of this model were done using SHELXL-97 (WinGX version, release 97-2) [15].

Crystal alignments were done using back-reflection Laue diffraction. Selected area electron diffraction was done using a Phillips CM300 atomic resolution transmission electron microscope with a field emission gun and a bottom mounted Orius CCD camera. The accelerating voltage was 300 kV.

**2.4 Magnetization and Specific Heat Measurements**

Magnetization and heat capacity measurements were made on both $FeSc_2S_4$ powder and single crystals using a Quantum Design Physical Properties Measurement System. Magnetization data was collected from $T = 1.9$ -300 K under applied fields of $\mu_0 H = 0.1$-0.5 T. For single crystals, the field was applied along the [100] direction. Zero field heat capacity was collected from $T = 1.9$ to $T = 70$ K. Field-dependent specific heat was collected up to $\mu_0 H = 14$ T from $T = 1.9$ to T = 20 *K*. Curie-Weiss analysis was performed over the temperature range 50 K $< T <$ 200 K, where previous reports indicate a region of linearity in the inverse susceptibility [1].

## 3. Results and Discussion

Rietveld refinements to a typical powder X-ray diffraction pattern from the as-prepared polycrystalline material, Fig. 1b, show a structure consistent with previous reports: cubic, spacegroup $Fd\bar{3}m$, with lattice parameter $a$ =10.5184(1) Å at room temperature. Previously reported values of *a* at room temperature are 10.606 Å [4] and 10.52 Å [5]. Further, no structural distortions other than minor lattice contraction were observed by laboratory powder X-ray diffraction down to 12 K (data not shown).

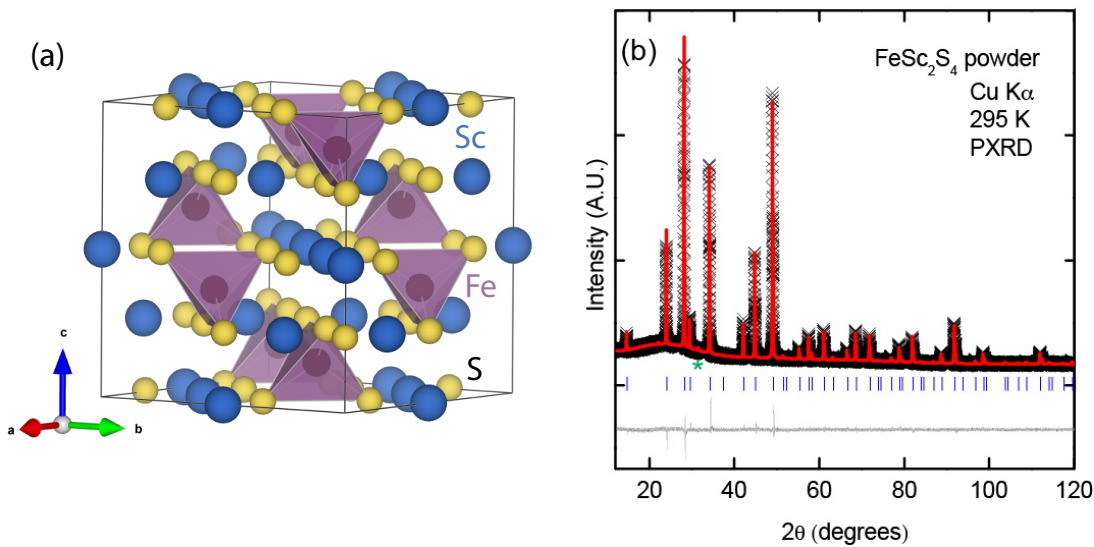

**Figure 1.** Structure and refinement of FeSc$_2$S$_4$ in spacegroup $Fd\bar{3}m$. (a) The spinel (AB$_2$X$_4$) unit cell of FeSc$_2$S$_4$. Fe (purple) is tetrahedrally coordinated by S (yellow) and sits on the A-site diamond sublattice. Sc (blue) is octahedrally coordinated and occupies the B-site. (b) Powder X-ray diffraction pattern of polycrystalline FeSc$_2$S$_4$. The experimental data is plotted as black symbols. A refinement of the model to the $Fd\bar{3}m$ space group is plotted as a red curve, and the difference between the data and the fit is plotted below in gray ($R_{wp}$ = 3.344). The *hkl* indices are represented by vertical ticks. The peak corresponding to an added internal Si standard is marked with a green asterisk.

As reported in the Sc$_2$S$_3$-FeS phase diagram [16], FeSc$_2$S$_4$ cannot be synthesized by simply melting and cooling of the stoichiometric composition since it does not melt congruently. However, this compound can be grown with excess FeS by the traveling solvent technique using a high temperature optical furnace. Fig. 2a shows a schematic of the mounting used to achieve this goal; given the high temperature (above 1517 °C) required, it was necessary to use a pyrolytic boron nitride crucible to avoid reaction with quartz; the boron nitride also acts as a thermal standoff to avoid melting the quartz.

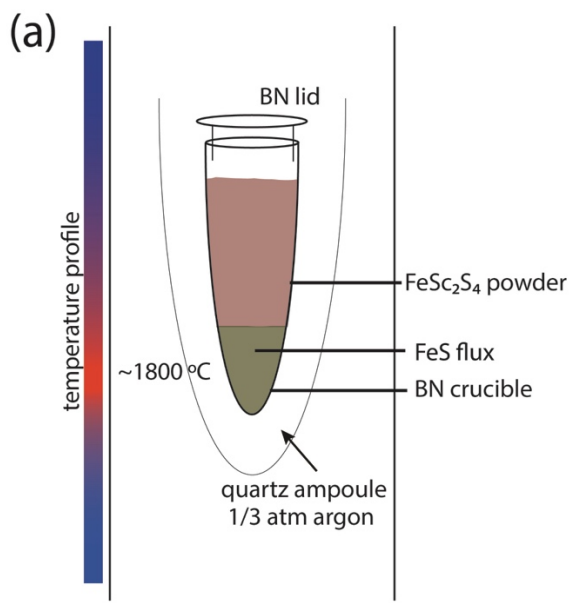

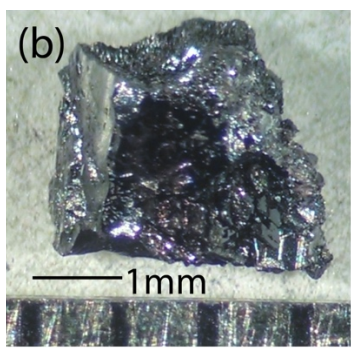

**Figure 2.** Schematic and example FeSc$_2$S$_4$ crystal. (a) Diagram of traveling solvent in a container using optical heating. The temperature profile used during the growth is indicated by a bar on the side, where the temperature of the hot zone (red) is above 1517 °C, based on the lamp power used and the phase diagram of this system [16]. (b) One crystal grown by this method, with dimensions approximately 4x4x2 mm. Facets are observable in this crystal.

A representative crystal grown by this technique is shown in Fig. 2b, oriented with the (100) plane facing the camera, confirmed by back-reflection Laue diffraction, Fig. 3a. Consecutive Laue patterns collected across the length of this crystal show a uniform orientation. A Laue pattern from a second crystal growth, also along the (100) plane, is shown in Fig. 3b.

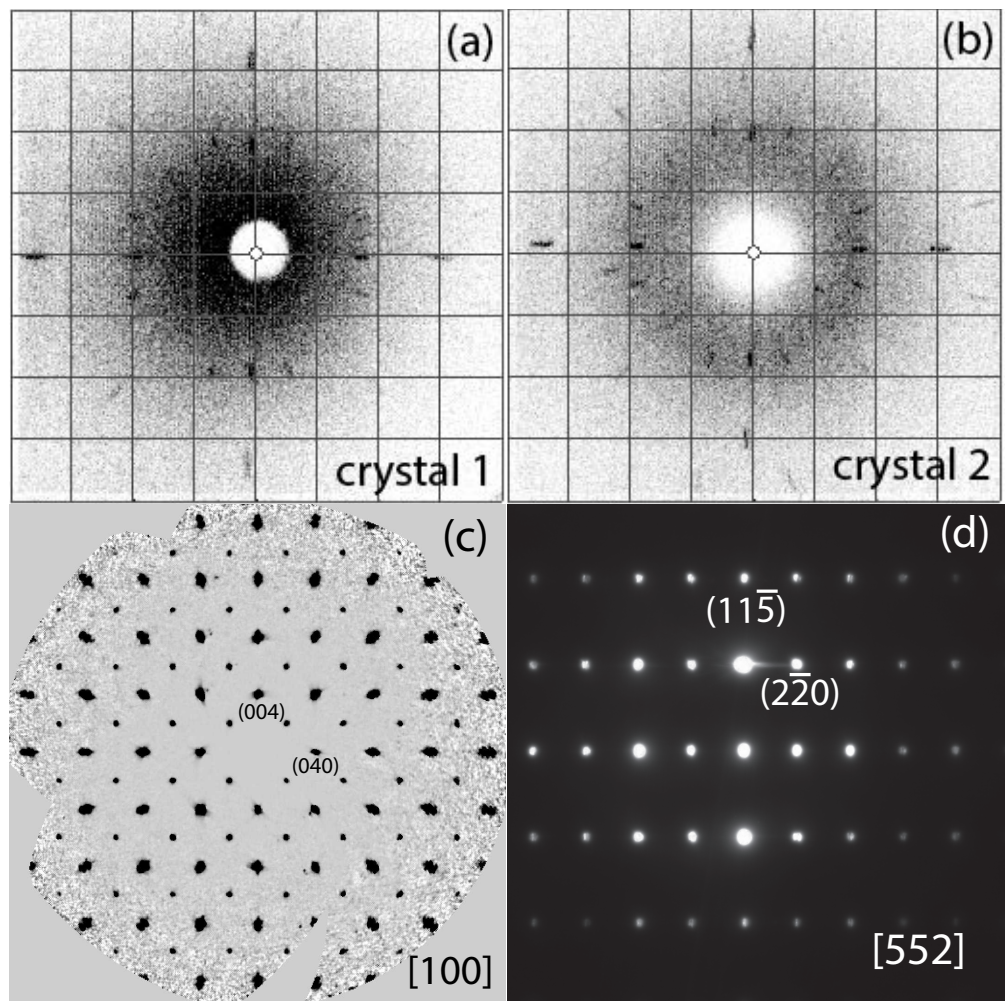

**Figure 3.** Diffraction patterns of single crystals and crystallites. (a-b) Representative back-reflection Laue diffraction patterns of the (100) face from $FeSc_2S_4$ crystals from two different growth runs. (c) Precession image from single crystal diffraction of $FeSc_2S_4$ crystal cut from a larger crystal piece. This image corresponds to the single crystal refinement results shown in Tables 1 and 2. (d) Selected area electron diffraction in the [552] direction from a crystallite in the polycrystalline sample. No diffuse scattering was observed.

To determine whether samples produced in this fashion have appropriate stoichiometry, we carried out an extensive series of structural and physical property measurements. First, as expected from the precession images, shown in Fig. 3c, modeling of single crystal X-ray diffraction data yielded a fit consistent with the $Fd\bar{3}m$ spacegroup, with a cubic lattice parameter $a = 10.5097(2)$ Å at $T = 110(2)$ K. This is also in agreement with selected area electron diffraction (SAED) images (Fig. 3d) of $FeSc_2S_4$, which show no diffuse scattering and indicate that at room temperature, $Fd\bar{3}m$ symmetry is maintained even on the local scale. Sulfur deficiency and partial occupancy of Fe or Sc sites, equivalent to a change in Fe:Sc ratio, were not observed in any tested models. Antisite mixing of the form $Fe_{1\pm x}Sc_{2\pm x}S_4$ did not improve the quality of the refinement, and were held fixed at their ideal values. Explicitly, 3.5% site mixing had the effect of increasing $R_1$ by 7% and $GooF$ by 10%. Thus, within the limits of detection of single crystal diffraction, the pieces produced here have appropriate stoichiometry.

**Table 1.** Crystallographic parameters for the first $FeSc_2S_4$ crystal obtained from model fits to the X-ray diffraction data. Absorption correction was analytical using a multifaceted crystal model.

| | |
|---|---|
| Temperature (K) | 110(2) |
| Space Group | $Fd\bar{3}m$ |
| $a$ (Å) | 10.5097(2) |
| $V$ (Å$^3$) | 1161 |
| Crystal Size (mm) | 0.155 x 0.146 x 0.127 |
| Collected Reflections | 7932/165 unique |
| $\theta_{max}$ / Completeness | 36.13 / 1.000 |
| $\mu$/mm | 6.022 |
| Transmission min/max | 0.503/0.576 |
| $R_{eq}$ | 0.0245 |
| $GooF$ | 1.417 |
| $R_1$ [$F^2 > 2\sigma(F^2)$] | 0.0261 |
| $wR_2$ ($F^2$) | 0.0686 |
| $\Delta\rho_{max}$ (Å$^{-3}$) | 2.59 |
| $\Delta\rho_{min}$ (Å$^{-3}$) | -1.42 |

**Table 2.** Atomic coordinates and atomic displacement parameters for $FeSc_2S_4$ in the $Fd\bar{3}m$ spacegroup. All occupancies refined to unity within error and thus were fixed at full occupancy in the final refinement.

| Atom | Wyckoff site | $x$ | $y$ | $z$ | $U_{11}=U_{22}=U_{33}$ (Å$^2$) |
|---|---|---|---|---|---|
| Fe1 | 8a | 1/8 | 1/8 | 1/8 | 0.0054(2) |
| Sc1 | 16d | 1/2 | 1/2 | 1/2 | 0.0048(2) |
| S1 | 32e | 0.25553(3) | 0.25553(3) | 0.25553(3) | 0.0049(2) |

A comparison between the physical properties of polycrystalline and single crystal samples was done to confirm the quality of the grown crystals of $FeSc_2S_4$. Fig. 4 shows a comparison of dc magnetic susceptibilities ($\chi \sim M/H$) at $T < 50$ K between previous literature powder data, data from the polycrystalline material prepared in this work, and data from two of our single crystals. Our powder and single crystals have similar behavior to the literature specimens, except in the low temperature regime where there is a pronounced roll-over in our measurements. The difference likely arises due to a greater density of defects in the literature samples: it is commonly observed in quantum magnetic materials that the presence of small non-stoichiometries or other defects results in the appearance of a Curie-tail, i.e. rising magnetic susceptibility, at low temperatures. This hypothesis is further supported by comparison of our data to literature $^{45}$Sc NMR data [3], which also shows a roll-over; NMR, unlike bulk magnetization measurements, directly probes the intrinsic local susceptibility and is much less sensitive to the defects that give rise to the Curie-tail effect. Based on these magnetization measurements, it appears that the polycrystalline samples here have a lower defect density than previously reported, and that the single crystals behave similarly.

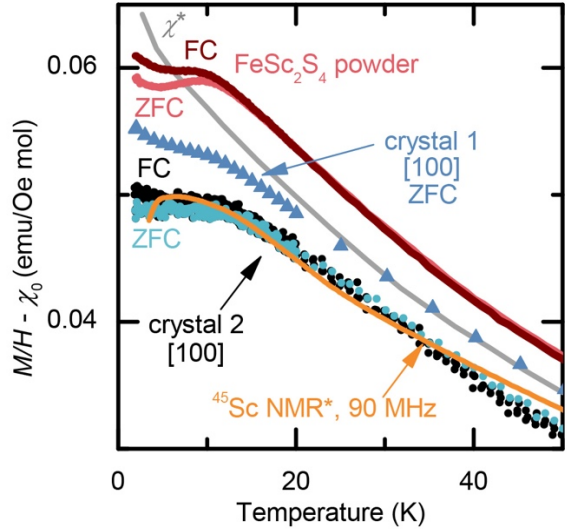

**Figure 4.** Magnetic susceptibility of polycrystalline and single crystal $FeSc_2S_4$. Comparison of magnetic susceptibilities of polycrystalline $FeSc_2S_4$ measured at $\mu_0H = 1T$ (solid gray) [1], the polycrystalline material in this work, measured at $\mu_0H = 0.5T$ (red points), two $FeSc_2S_4$ crystals oriented to [100] at $\mu_0H = 0.1T$ (light blue/black circles) and $\mu_0H = 0.5T$ (dark blue triangles) and the susceptibility measured via $^{45}$Sc NMR Knight Shift at 90 MHz (solid orange line) [3]. Powder and crystal measurements from this work exhibit a peak at $T = 11K$, in agreement with the NMR Knight shift data. Asterisks denote data taken from literature.

A Curie-Weiss analysis of the magnetic data in the paramagnetic regime gives parameters in agreement with previous reports, Table 3. The Weiss temperatures, $\theta_w$, are negative, indicating net mean field antiferromagnetic interactions. The Curie constants, $C$, correspond to effective magnetic moments, $p_{eff}$, that are in good agreement with the spin-orbital model for $Fe^{2+}$ on a tetrahedron that predicts $p_{eff} = 5.32$. One significant difference between our results and prior literature is the presence of a non-negligible temperature-independent contribution, $\chi_0$, in the present samples. This is most likely due to the presence of residual contributions from residual travelling solvent, FeS, which is a ferromagnetic metal at all temperatures measured here. Residual FeS is likely contributing to the unreliably large $\theta_w$ in Crystal 2 (Table 3), which is significantly more negative than is seen in the other samples. The error on this calculated value is particularly large (20%), but the calculated $\chi_0$-subtracted susceptibility is well in line with other samples in this and other work (Figure 4). Based on the reported saturation magnetization for FeS, we can estimate between 0.05% and 0.2% FeS in samples reported here [17]. The presence of these inclusions precludes more precise comparisons. Future optimization of the traveling solvent technique as applied here is expected to be able to eliminate these inclusions.

**Table 3.** Curie-Weiss analysis of $FeSc_2S_4$: polycrystalline material in this work and reported in literature [1], and two grown crystals. $C$ is the Curie constant (emu K $Oe^{-1}$ mol f.u.$^{-1}$), $\theta_w$ (K) the Weiss Temperature, $p_{eff}$ the effective magnetic moment per ion, and $\chi_0$ the temperature independent contribution to the magnetic susceptibility (emu $Oe^{-1}$).

|  | Powder | Lit. Powder | Crystal 1 | Crystal 2 |
|---|---|---|---|---|
| $C$ | 3.45(5) | 3.28 | 3.60(5) | 3.5(2) |
| $\theta_w$ | -42(2) | -45.1 | -54(5) | -100(20) |
| $p_{eff}$ | 5.3(1) | 5.12 | 5.4(3) | 5.3(3) |
| $\chi_0$ | 0.0311 |  | 0.07274 | 0.17748 |

Fig. 5 shows a comparison of heat capacity measurements on literature powder versus the polycrystalline and single crystal samples prepared in this work. In all cases, there is a broad maximum in $C/T$ at $T \sim 10$ K. The maximum is at a slightly higher temperature and sharper in the specimens of this work, consistent with the presence of fewer defects (which tend to broaden transitions). These differences are not attributable to the presence of FeS inclusions, since there is only a weak field dependence to the specific heat (inset). The magnetic contribution to specific heat ($C_{magnetic}$) was estimated by subtracting the phononic contribution ($C_{phonon}$) from the total specific heat in each case. We used existing literature data on $CdIn_2S_4$, which has no magnetic degrees of freedom, and scaled it based on the change in atomic masses per known methods [18]. The results are shown in Fig. 5 and are similar amongst all four datasets. Note that our integrated entropies differ from that of Fritsch et al. [1] due to a difference in how the non-magnetic heat capacity was scaled (future work with a more closely atomic weight matched analog is necessary to unambiguously resolve which is more accurate). Both the powder and the single crystals recover entropy to approximately 70% of Rln(5). The value Rln(5) corresponds to the spin and orbital degrees of freedom for $Fe^{2+}$ in a tetrahedral coordination. The presence of a transition at low temperatures was explored by Plumb et al., [7] who observed evidence for a tetragonal distortion in this material at $T \sim 11K$. The physical basis of this peak at low temperatures is still under investigation. Synthesis of large single crystals will further enable this exploration.

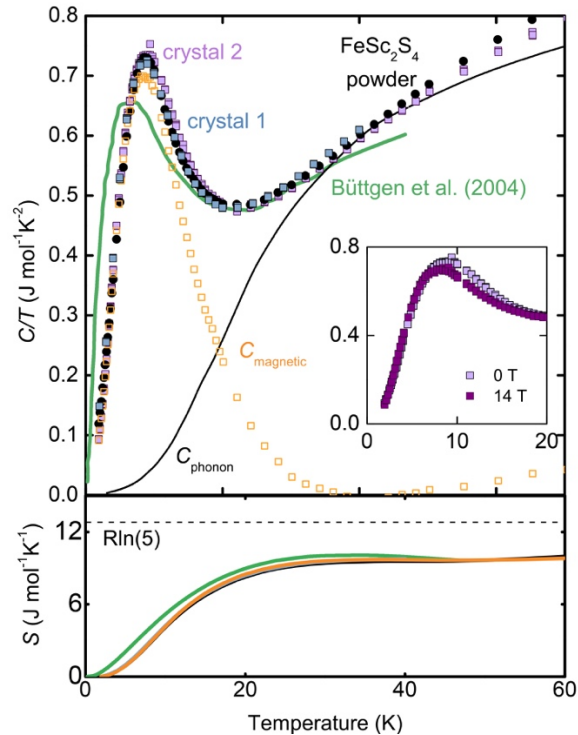

**Figure 5.** Heat capacity of polycrystalline and single crystal $FeSc_2S_4$. (a) Heat capacity measurements on crystal 1 (blue squares) and crystal 2 (purple squares) are consistent with measurements on the powder (black circles). Both show a broad peak between 2-15 K that responds minimally to fields up to 14 T. The peak is slightly higher in temperature and sharper than reported heat capacity data for polycrystalline samples [1]. (b) Integrated entropy, and thus number of spin and orbital degrees of freedom involved, are similar between all four samples.

## 4. Conclusion

In short, we report the successful preparation of polycrystalline powder and mm-scale single crystals of stoichiometric iron scandium sulfide by the travelling solvent technique. This paves the way to finally understanding this fascinating material, with future optimizations to improve the purity and size of single crystals. The use of an optical furnace to perform a traveling solvent crystal growth in a container is also adaptable to a wide range of other complex quantum materials [13], and growths of other materials by related techniques are in progress.

*Corresponding Authors, e-mail: mcqueen@jhu.edu, koohpayeh@jhu.edu

Shared mailing address: 3400 N. Charles St, Department of Chemistry, The Johns Hopkins University, Baltimore, MD, 21218

## Acknowledgements

The authors would like to thank W. Adam Phelan for assisting with single crystal diffraction refinements and Collin L. Broholm for useful discussions and support. This research was supported by the US Department of Energy, Office of Basic Energy Sciences, Division of Materials Sciences and Engineering under Award DE-FG-02-08ER46544 to the Institute for Quantum Matter at JHU. TMM acknowledges support from the David and Lucile Packard Foundation and the Sloan Research Fellowship.